\documentclass{elsart}

\usepackage{graphicx}
\usepackage{amssymb}
\usepackage{lineno}

\begin{document}

\begin{frontmatter}
\title{Scintillation-only Based Pulse Shape Discrimination for Nuclear and Electron Recoils in Liquid Xenon}
\author{K.Ueshima\thanksref{label1}\corauthref{cor1}}, 
\ead{ueshima@suketto.icrr.u-tokyo.ac.jp}
\corauth[cor1]{Corresponding Author}
\author{K.Abe\thanksref{label1}},
\author{K.Hiraide\thanksref{label1}},
\author{S.Hirano\thanksref{label1}},
\author{Y.Kishimoto\thanksref{label1}\thanksref{label22}},
\author{K.Kobayashi\thanksref{label1}},
\author{Y.Koshio\thanksref{label1}},
\author{J.Liu\thanksref{label22}},
\author{K.Martens\thanksref{label22}},
\author{S.Moriyama\thanksref{label1}\thanksref{label22}},
\author{M.Nakahata\thanksref{label1}\thanksref{label22}},
\author{H.Nishiie\thanksref{label1}},
\author{H.Ogawa\thanksref{label1}},
\author{H.Sekiya\thanksref{label1}},
\author{A.Shinozaki\thanksref{label1}},
\author{Y.Suzuki\thanksref{label1}\thanksref{label22}},
\author{A.Takeda\thanksref{label1}},
\author{M.Yamashita\thanksref{label1}\thanksref{label22}},
\author{K.Fujii\thanksref{label3}},
\author{I.Murayama\thanksref{label3}},
\author{S.Nakamura\thanksref{label3}},
\author{K.Otsuka\thanksref{label2}},
\author{Y.Takeuchi\thanksref{label2}\thanksref{label22}},
\author{Y.Fukuda\thanksref{label4}},
\author{K.Nishijima\thanksref{label5}},
\author{D.Motoki\thanksref{label5}},
\author{Y.Itow\thanksref{label6}},
\author{K.Masuda\thanksref{label6}},
\author{Y.Nishitani\thanksref{label6}},
\author{H.Uchida\thanksref{label6}},
\author{S.Tasaka\thanksref{label8}},
\author{H.Ohsumi\thanksref{label23}},
\author{Y.D.Kim\thanksref{label10}},
\author{Y.H.Kim\thanksref{label9}},
\author{K.B.Lee\thanksref{label9}},
\author{M.K.Lee\thanksref{label9}},
and
\author{J.S.Lee\thanksref{label9}}\\
{the XMASS Collaboration}
\address[label1]{Kamioka Observatory, Institute for Cosmic Ray Research, The University of Tokyo, Kamioka, Hida, Gifu 506-1205, Japan}

\address[label22]{Institute for the Physics and Mathematics of the Universe, The University of Tokyo, Kashiwa, Chiba 277-8582, Japan}

\address[label3]{Department of Physics, Faculty of Engineering, Yokohama National University, Yokohama 240-8501, Japan}

\address[label2]{Department of Physics, Kobe University, Kobe, Hyogo 657-8501, Japan }

\address[label4]{Department of Physics, Miyagi University of Education, Sendai, Miyagi 980-0845, Japan}

\address[label5]{Department of Physics, Tokai University, Hiratsuka, Kanagawa 259-1292, Japan}

\address[label6]{Solar Terrestrial Environment Laboratory, Nagoya University, Nagoya, Aichi 464-8602, Japan }

\address[label8]{Department of Physics, Gifu University, Gifu, Gifu 501-1193, Japan}

\address[label23]{Faculty of Culture and Education, Saga University, Honjo, Saga 840-8502, Japan}

\address[label10]{Department of Physics, Sejong University, Seoul 143-747, Korea}
 
\address[label9]{Korea Research Institute of Standards and Science, Daejeon 305-340, Korea}

\begin{abstract}
In a dedicated test setup at the Kamioka Observatory we studied pulse shape 
discrimination (PSD) in liquid xenon (LXe) for dark matter searches. 
PSD in LXe was based on the observation that scintillation light from electron events was emitted over a longer period of time than that of nuclear 
recoil events, and our method used a simple ratio of early to total 
scintillation light emission in a single scintillation event.
Requiring an efficiency of 50\% for nuclear recoil retention 
we reduced the electron background to 
7.7$\pm$1.1(stat)$\pm ^{1.2}_{0.6}$(sys)$\times$10$^{-2}$
at energies between 4.8 and 7.2\,keV$_{ee}$ and to 
7.7$\pm$2.8(stat)$\pm ^{2.5}_{2.8}$(sys)$\times$10$^{-3}$
at energies between 9.6 and 12\,keV$_{ee}$ for a scintillation light yield of 
20.9\,p.e./keV. 
Further study was done by masking some of that light to reduce this yield to 4.6\,p.e./keV, the same 
method results in an electron event reduction of 
2.4$\pm$0.2(stat)$\pm ^{0.3}_{0.2}$(sys)$\times$10$^{-1}$
for the lower of the energy regions above. 
We also observe that in contrast to nuclear recoils the fluctuations in our 
early to total ratio for electron events are larger than expected from 
statistical fluctuations. 

%% Pulse shape discrimination (PSD) of liquid xenon for dark matter search is studied using a detector with high scintillation light yield.
%% The rejection power, the reduction of electron events keeping the detection efficiency of nuclear recoil events at 50\%, is 7.7$\pm$1.1(stat)$\pm ^{1.2}_{0.6}$(sys)$\times$10$^{-2}$ in the energy range between 4.8-7.2 keV$_{ee}$ and 7.7$\pm$2.8(stat)$\pm ^{2.5}_{2.8}$(sys)$\times$10$^{-3}$ in the energy range between 9.6-12 keV$_{ee}$ at high light yield (20.9 p.e./keV). 

%%  The light yield dependence of the rejection power was also studied. The rejection power at low light yield (4.6 p.e./keV) was 2.4$\pm$0.2(stat)$\pm
%%  ^{0.3}_{0.2}$(sys)$\times$10$^{-1}$ in the energy range between 4.8-7.2 keV$_{ee}$.

%%  In addition the scintillation waveform for electron events is observed to have a larger spread than that estimated by the photo-electron statistics.  

\end{abstract}

\end{frontmatter}

%\linenumbers
% insert the table of contents
\pagenumbering{roman}
%\tableofcontents
% <Chapter>
%\newpage
\pagenumbering{arabic}
\pagestyle{plain}

\section{Introduction}
 
 The results of various astronomical observations \cite{be} - \cite{sdss} show 
strong evidence for a large amount of dark matter in the universe. 
Weakly Interactive Massive Particles
(WIMPs) are a dark matter candidate motivated in extensions of the theory of 
high energy particle physics \cite{jung}. 
Various dedicated WIMP dark matter search experiments are underway around the 
world \cite{dama} - \cite{xe10}. The XMASS experiment, using liquid xenon 
(LXe) as a target for WIMP dark matter, was proposed in 2000 \cite{xmass}. 
The construction of the 800\,kg detector was finished in 2010. 

 The interaction of WIMP dark matter is observed as a nuclear recoil in a 
specific detector's target material, which in our case is LXe. 
The main backgrounds (BG) for such nuclear recoil events are electron events (from photoabsorption or Compton scattering of environmental gamma rays), 
nuclear recoils from the scattering of fast neutrons, 
and possibly alpha and beta decays in the detector medium itself. 
The aim of this study is to use the shape of scintillation light pulse in LXe  to discriminate against electron events. 

   The scintillation mechanism is classified into two processes determined by whether or not the process includes recombination \cite{emi}.   
 
\textbf{the process without recombination }
\begin{eqnarray}
\textrm{Xe}^{*} + \textrm{Xe} & \rightarrow & \textrm{Xe}^{*}_{2} \nonumber \\
\textrm{Xe}^{*}_{2} & \rightarrow & 2\ \textrm{Xe} + h \nu 
\label{eq:fast}
\end{eqnarray}

\textbf{the process with recombination}
\begin{eqnarray}
\textrm{Xe}^{+} + \textrm{Xe} & \rightarrow & \textrm{Xe}^{+}_{2} \nonumber \\
\textrm{Xe}^{+}_{2} + e^{-} & \rightarrow & \textrm{Xe}^{**} + \textrm{Xe} \nonumber \\
\textrm{Xe}^{**} & \rightarrow & \textrm{Xe}^{*} + \textrm{heat} \nonumber \\
\textrm{Xe}^{*} + \textrm{Xe} & \rightarrow & \textrm{Xe} ^{*} _{2} \nonumber \\
\textrm{Xe}^{*}_{2} & \rightarrow & 2\ \textrm{Xe} + h \nu 
\label{eq:slow}
\end{eqnarray}
 
  Both processes lead to the formation of exciton ( $\textrm{Xe}^{*}_{2}$ ). The de-excitation has two components called singlet and triplet component. The singlet component is caused by a spin singlet state( $^{1}\Sigma_{u}^{+}$ ), and the triplet component is caused by a spin triplet state( $^{3}\Sigma_{u}^{+}$ ) \cite{x10}.  

 The shape of the scintillation light pulse in LXe is determined by the lifetimes of the excited states, the time scale of electron-ion recombination, the 
timing resolution of the photo detector, the time of flight in the detector, and electronics employed to record the 
scintillation light. The convolution of all these components shapes the 
recorded scintillation light pulses. 
 
 Pulse shape measurements for 1\,MeV electrons, $\alpha$ particles and 
relativistic ions in LXe were reported \cite{x5}. In the case of $\alpha$ 
particles, two distinct lifetime components were observed. The lifetimes for 
singlet and triplet states were found to be approximately 4\,ns and 22\,ns
respectively. 
In the case of electron events on the other hand, only one component with 45\,ns lifetime was observed.

 This pulse shape difference is attributed to the  influence of electron-ion recombination \cite{x14} \cite{x18}. 
In the case of electron event, dE/dx is one order of magnitude smaller than that in the case of a nuclear recoil \cite{x14} \cite{rec}, and the ensuing 
ionization is thus spread out over a larger volume. 
Therefore the recombination process of electron events takes longer 
than that of nuclear recoil events and dominates the pulse evolution. 
This hypothesis was confirmed by measurements in electric fields, as the slow 
component was not observed when an external electric field is applied. 
Under the influence of an external electric field recombination is suppressed 
by the drifting apart of the opposing charges, and the two components 
characteristic of high dE/dx events re-emerge, so that 
the pulse shape of electron events in an electric field of 4\,kV/cm is very close to that of nuclear recoils in zero electric field \cite{x10}. 
In this paper we aim to exploit the characteristically long time constant 
observed for electron events in LXe at zero electric field. 

To this end we examine pulse shape discrimination (PSD) in 
LXe at energies of less than 20\,keV$_{ee}$ ( electron equivalent keV). This 
energy range is most relevant for dark matter searches. The average waveform 
of nuclear recoil and electron events was previously compared above 
10\,keV$_{ee}$ \cite{x20}. But no event-by-event analysis was done. An event-by-event analysis was reported \cite{x19}, but the light yield was one order of magnitude lower than our setup.
Using dual phase detectors the PSD study was also performed \cite{dis3}. A reduction factor for electron events of 0.2 by PSD alone 
was demonstrated at 5\,keV$_{ee}$ in a dual phase detector by limiting 
the electric field to 0.06\,kV/cm.

Previous investigations (\cite{x20} - \cite{dis3}) were done at scintillation 
light yields below 5\,p.e./keV (p.e. photoelectrons). Using two closely spaced 
photomultipliers (PMTs) of the kind also used in the XMASS experiment we 
measure a light yield of 20.9\,p.e./keV, which is expected to us to clearly observe a 
difference in the pulse shapes of nuclear and electron events even at 
energies as low as 5\,keV$_{ee}$. Anticipating more limited light collection 
in real detectors we artificially reduced the photosensitive area in our 
experimental setup and repeated the measurements at an effective light 
yield of 4.6\,p.e./keV.

\section{Detector setup}

\begin{figure}[h]
\begin{center}
\includegraphics[scale = 0.3]{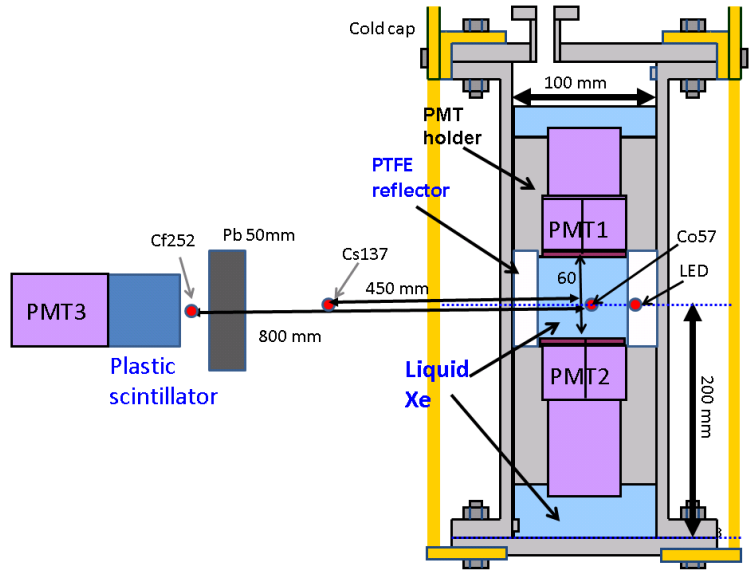}
\caption{A schematic view of the detector used in the PSD measurement.}
\label{fig:6_1}
\end{center}
\end{figure}

 Our measurements were made in a dedicated setup shown in Fig.\ref{fig:6_1}. 
Two 2\,inch hexagonal PMTs (Hamamatsu R10789) facing each other from a distance of 6\,cm are viewing 
the intervening LXe volume from the top and bottom respectively. To the sides 
the volume is limited by a highly reflective PTFE surface enclosing 0.58\,kg of LXe in total. Embedded in the PTFE reflector were a LED and a 300\,Bq 
$^{57}$Co source. The LED was used to obtain single p.e. spectra and the 
source to monitor the PMT gain. 
The light yield from this source was found to be stable within 2\% and is 
incorporated in our systematic uncertainty. 
To protect the LXe from radioactive contamination this $^{57}$Co source 
was enclosed in a thin stainless steel container. 

 PMTs, LXe, and the calibration light sources described above were all kept at 
-100$^{\circ}$C inside a stainless steel vessel that 
itself is suspended inside a vacuum chamber for thermal insulation. 
Radioactive sources placed outside of the LXe volume and its vacuum enclosure 
produce the recoil events used 
in this study: Either a $^{137}$Cs source provides gamma rays that produce 
electron events or a gamma tagged $^{252}$Cf fission source provides 
neutrons to study nuclear recoils. To tag the fission events a plastic scintillator and a PMT (PMT3) were set up next to the Cf source. In the direction of the LXe on the other hand the gamma 
rays emitted in the $^{252}$Cf fission events were shielded by 5\,cm of lead. 

 When filling the setup with LXe the Xe gas was passed through a SAES getter (Model PS4-MT3-R-1) to remove impurities in the Xe gas. Beyond this no further efforts 
were made to clean up the high purity Xe used in this experiment. 

\begin{figure}[h]
\begin{center}
\includegraphics[scale = 0.7]{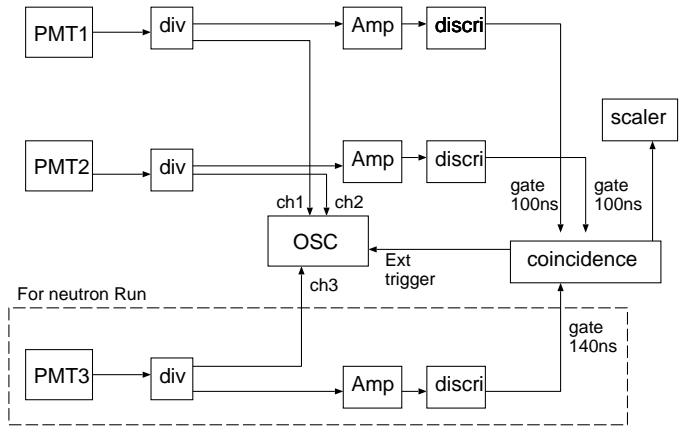}
\caption{A schematic diagram of the data acquisition system.}
\label{fig:trig}
\end{center}
\end{figure}

 For data acquisition we used NIM logic to trigger recording of the 
waveform in a LeCroy WavePro 900 digital oscilloscope as shown in 
Fig.\ref{fig:trig}. 
Events in the LXe volume are identified by a coincidence of signals in the 
two PMTs that view the LXe volume. 
The width of the coincidence timing window is 100\,ns, and the discriminator 
thresholds were set to 2\,p.e. equivalent. 
As mentioned above a plastic scintillator is used to 
tag $^{252}$Cf fission events every time the Cf source is employed to produce 
neutron events in the LXe. 
The corresponding signal from the PMT that reads out the plastic scintillator 
starts a 140\,ns gate that is only for neutron recoil data is also entered 
into the coincidence unit. 

 The digital oscilloscope provides 8\,bit resolution and a 1\,GHz sampling rate. 
We used it to record 10\,$\mu$s traces of all PMT signals involved in 
the respective measurement with 1\,ns timing resolution. 

\begin{figure}[h]
\begin{center}
\includegraphics[scale = 0.5]{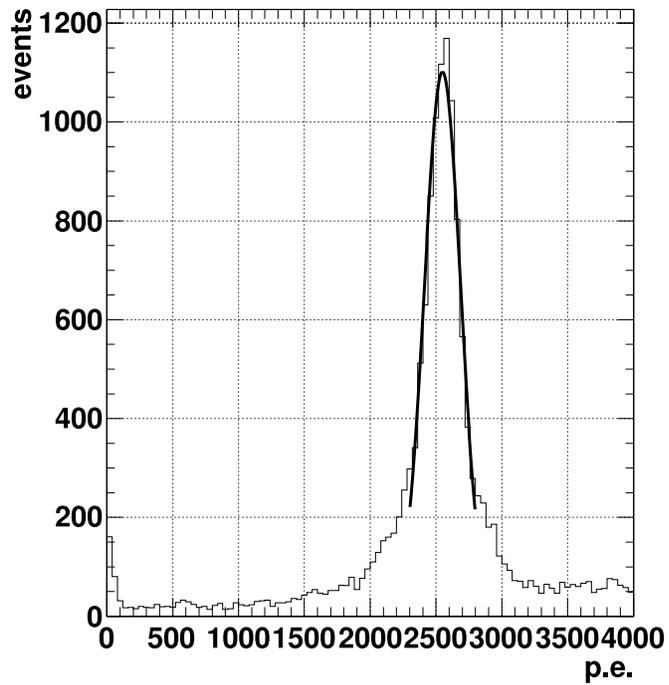}
\caption{A typical total p.e. distribution of $^{57}$Co data.}
\label{fig:co}
\end{center}
\end{figure}

 The gain of the two PMTs reading out the LXe volume were both set to 
3.8$\times$10$^{6}$ at the operating temperature of -100\,$^{\circ}$C, with the 
HV for PMT1 and PMT2 fixed at 1.26\,kV and 1.21\,kV, respectively. 
Using the 1\,p.e. calibration obtained from the LED data, the fit shown in 
Fig.\ref{fig:co} to the $^{57}$Co spectrum determines the energy resolution to 
be 5.4\% (RMS) at 122\,keV and the scintillation light yield to be 
20.9\,p.e./keV$_{ee}$. 

Light collection in real dark matter detectors will be more limited than in 
this small scale dedicated setup. To quantify the impact of the ensuing loss 
in statistical power on our discrimination method we artificially reduced the 
light collection in our setup by covering part of the photocathode area of the 
bottom PMT (PMT2) with a copper mask. While the PMT1 signal was no longer used to estimate the deposited energy in our study of reduced effective light yield, its signal was used in the trigger during that study in just same way as usual.

\begin{figure}[h]
\begin{center}
\includegraphics[scale = 0.5]{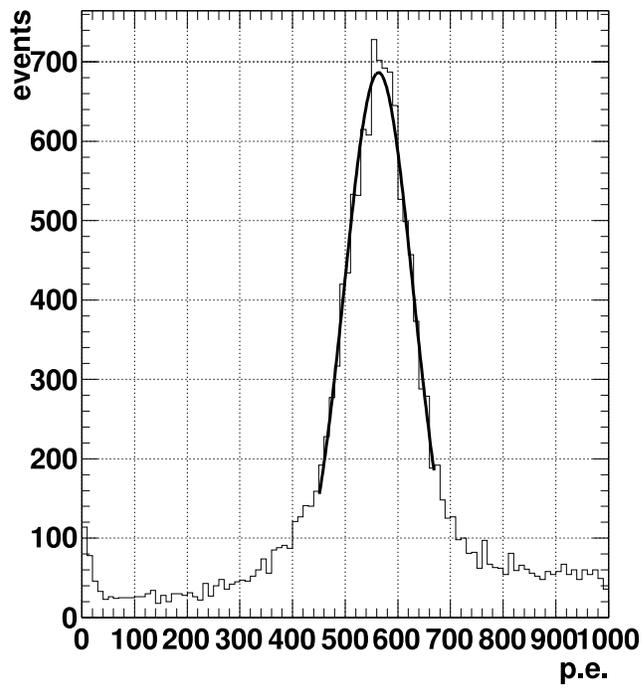}
\caption{A typical p.e. distribution of $^{57}$Co data for PMT2. The light yield was reduced to 4.6\,p.e./keV$_{ee}$ by a mask.}
\label{fig:pmt2_mask_co}
\end{center}
\end{figure}

 The resulting $^{57}$Co spectrum is shown in 
Fig.\ref{fig:pmt2_mask_co}, using the p.e. count of only this bottom PMT. 
The effective light yield was thus reduced to 4.6\,p.e./keV$_{ee}$ and the 
energy resolution to 11.7\% (RMS) at 122 keV. 

\section{Data reduction}

 A trigger offset of 260\,ns allowed us to monitor the baselines 260\,ns prior 
to the recording of our physics events. 
To clean up both our electron and nuclear recoil event samples we require that 
none of the events show any ``pre-activity'' in the first 150\,ns from the 
beginning of its PMT traces, i.e. from 260\,ns to 110\,ns prior to the event 
trigger time. 

 The dynamic range of oscilloscope was chosen to saturate at a signal height of -400\,mV for the 
PMTs that read out our LXe volume. To avoid problems with saturation, we 
disregard events where PMT traces from either PMT1 or PMT2 surpass 350\,mV, 
which is roughly equivalent to 1000\,p.e. for electron events. 

\begin{figure}[h]
\begin{center}
\includegraphics[scale = 0.5]{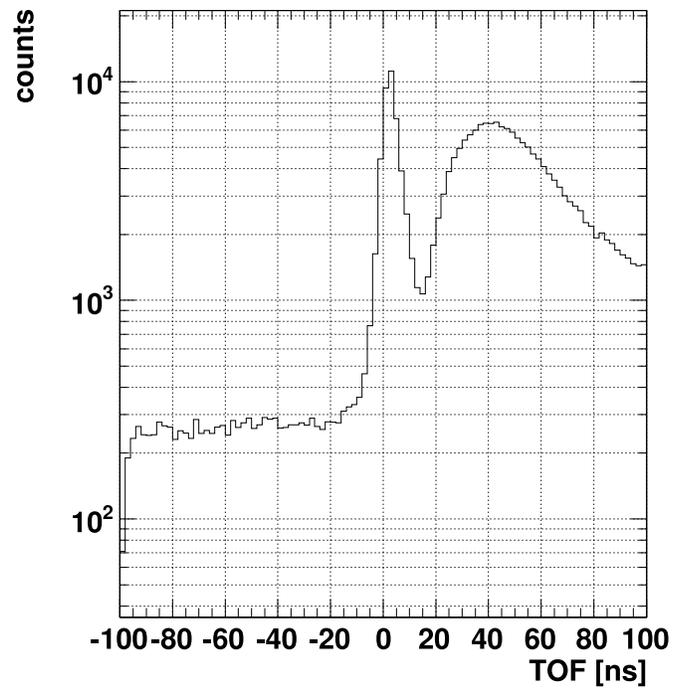}
\caption{TOF distribution of $^{252}$Cf Run.}
\label{fig:cf_tim}
\end{center}
\end{figure}

 Neutron events are identified by the neutron's time of flight (TOF) in the 
offline analysis. 
Fig.\ref{fig:cf_tim} shows the TOF distribution extracted from software  
threshold crossings between the leading edges of PMT3 and PMT2. 
The software thresholds for all timing determinations corresponds to the 
typical height of a 3\,p.e. signal from the PMTs. 
The peak at zero timing is due to $^{252}$Cf fission gammas passing through the 
lead shielding set up to suppress such direct gamma interference. 
To limit the gamma ray background in the sample and specifically select fast 
neutrons a narrow TOF range from 15\,ns to 30\,ns was chosen for further analysis. 

 Table \ref{tab:cut} summarizes our data reduction:
 
\begin{table}[h]
 \begin{center}
 \begin{tabular}{|c|c|c|}
 \hline
& $^{252}$Cf (neutron) Run & $^{137}$Cs (gamma ray) Run   \\
 \hline
before cuts & 8.0$\times 10^{5}$ &  4.0$\times 10^{5}$   \\
 \hline
after pre-activity cut & 5.6$\times 10^{5}$ (69\%)   
& 3.5$\times 10^{5}$ (88\%)  \\
 \hline
after saturation cut  & 2.2$\times 10^{5}$ (27\%)    
& 7.4$\times 10^{4}$ (18\%)   \\
 \hline
after TOF selection &  2.2$\times 10^{4}$ (2.7\%)   & \\
 \hline
 \end{tabular}
 \end{center}
 \caption{Data reduction summary of PSD measurement.}
 \label{tab:cut}
\end{table}
 
\section{Pulse shape discrimination}

 The parameter we chose to discriminate between electron and nuclear recoil 
events is the ratio $R_{PSD}$ of scintillation light detected in just the first 
$\Delta t_{t1}=$20\,ns of the scintillation light pulse, which we refer to as the 
prompt light, to the total amount of scintillation light detected in that same 
pulse. As we measure light with PMTs, the amounts of prompt and total light 
are recorded in units of p.e.: $p.e._{prmt}$ and $p.e._{tot}$.
The window for evaluating the amount of total light starts at the same time 
$t_0$ as that for the prompt light and is $\Delta t_{tot}=$200\,ns long.
$t_0$ is the same software threshold 
crossing time as used in the TOF distribution above. 
Therefore each PMT has its own software threshold crossing resulting in 
potentially slightly different start times $t_{0,1}$ and $t_{0,2}$ for the 
integration of a PMT's respective scintillation signal. 
The respective integrals for prompt and total charges measured from the 
base line subtracted 
oscilloscope traces $V_{PMT1}(t)$ and $V_{PMT2}(t)$ of the scintillation signal 
recorded by PMT1 and PMT2 respectively are added as the ratio is evaluated
for each event:

\begin{eqnarray}
 R_{PSD} = \frac{\int_{t_{0,1}}^{t_{0,1}+\Delta t_{t1}}V_{PMT1}\,dt 
+ \int_{t_{0,2}}^{t_{0,2}+\Delta t_{t1}}V_{PMT2}\,dt}
{\int_{t_{0,1}}^{t_{0,1}+\Delta t_{tot}}V_{PMT1}\,dt 
+ \int_{t_{0,2}}^{t_{0,2}+\Delta t_{tot}}V_{PMT2}\,dt} = \frac{p.e._{prmt}}{p.e._{tot}}
\end{eqnarray} 

The 20\,ns width for the prompt timing window was optimized for best 
discrimination against gamma recoils. In the case of the reduced effective 
light yield measurement all terms involving $V_{PMT2}$ are simply dropped. 
A possible systematic effect of the recoil events position in the LXe volume 
was studied with the help of the light balance in the two PMTs. 

\subsection{Neutron data} 

  The recoil energy $E_{r}$ of a xenon nucleus elastically scattered on by a 
neutron with energy $E_{n}$ is expressed by the following equation:
\begin{eqnarray}
E_{r} = E_{n} \frac{2(A+1-\cos^{2}\theta - 
\cos\theta \sqrt{A^{2}-1+\cos^{2}\theta})}{(1+A)^{2}}
\end{eqnarray} 
 where $A$ is the mass number of the recoil nucleus, and $\theta$ is the 
scattering angle of the neutron.
 The maximum recoil energy at $\theta \sim 180^{o}$ is 220\,keV$_{r}$ for an
8\,MeV neutron. 
To estimate the light yield we need to know the relative scintillation 
efficiency $L_{eff}$ that describes the scintillation light yield of a nuclear 
recoil event as compared to the yield of an electron event at the same energy; we assume $L_{eff}=0.2$ \cite{xe10}. 
With that the maximum visible energy for a Xe nucleus recoiling in LXe from the elastic scattering of an 8\,MeV neutron is 44\,keV$_{ee}$, 
corresponding to about 920\,p.e. in our setup.

 The trigger rate for the neutron data run with the $^{252}$Cf source was 
13.2\,Hz (coincidence rate of PMT1\&PMT2\&PMT3).
 The trigger rate of PMT3 alone was 16.2\,kHz and the coincidence of just 
PMT1 and PMT2 occurred at 1.12\,kHz. 
 
 Accidental coincidences in the sample can be estimated from the background in 
the TOF distributions for each energy range individually, and are found to be 
less than 5\% after our event selection. 

\begin{figure}[h]
\begin{center}
\includegraphics[scale = 0.5]{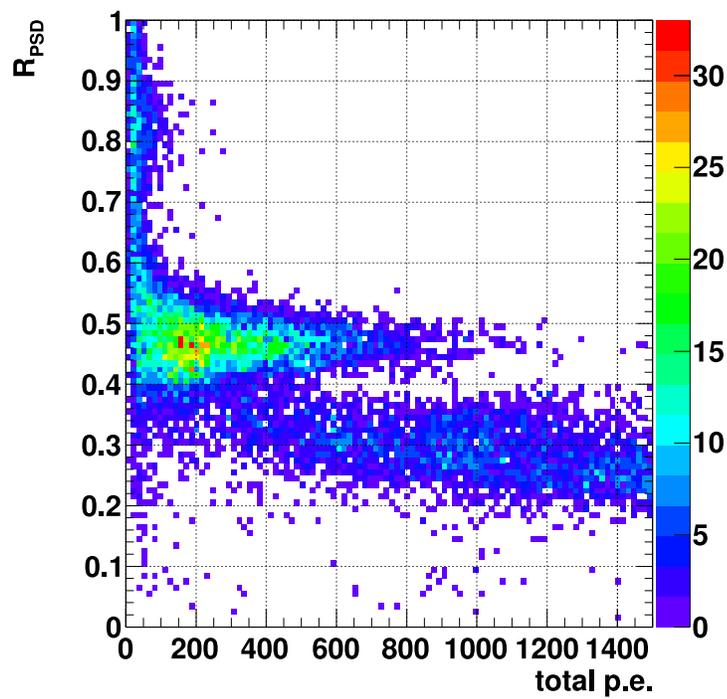}
\caption{ The scatter plot between total p.e. and $R_{PSD}$ for TOF from 15 to 30\,nsec in the $^{252}$Cf data. }
\label{fig:psd_cf1}
\end{center}
\end{figure}

 Fig.\ref{fig:psd_cf1} shows the correlation between total p.e. and $R_{PSD}$
for the neutron data. 
The nuclear recoil band just below $R_{PSD}=0.5$ can clearly be seen, ending 
about where expected from the above calculation of maximal recoil energy. 
At higher p.e. we also find gamma rays caused by inelastic scattering below $R_{PSD}=0.4$, which is well separated from the nuclear recoils.  
 The cluster at high $R_{PSD}$ $\sim$ 0.8 was caused by very sharp pulse such as Cherenkov light generated in the PMT window.

\subsection{Gamma-ray data}

\begin{figure}[h]
\begin{center}
\includegraphics[scale = 0.5]{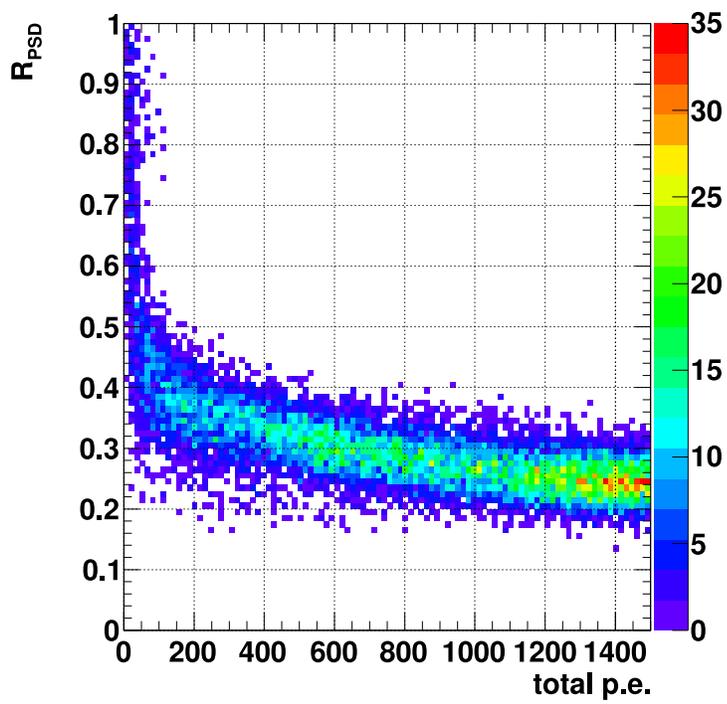}
\caption{The $^{137}$Cs scatter plot between total p.e. and $R_{PSD}$.}
\label{fig:psd_cs}
\end{center}
\end{figure}

Gamma ray data was taken with an external $^{137}$Cs source and a lead 
collimator for the source.
The trigger rate of the $^{137}$Cs Run was 4.52\,kHz (PMT1\&PMT2).
Fig.\ref{fig:psd_cs} shows the correlation between total p.e. and $R_{PSD}$ for 
this electron event sample. 
While the average $R_{PSD}$ was rather constant for nuclear recoils in the 
$^{252}$Cf data down to low energies, it does increase for gamma-rays at low 
energies, slowly merging into the nuclear recoil region as recoil energies 
approach zero. 

\section{Results}

 In Fig. \ref{fig:psd_cf1} and Fig. \ref{fig:psd_cs} we expect that power of 
$R_{PSD}$ to isolate nuclear recoils and reject gamma-ray background is a 
function of the scintillation light output observed in our recoil event. 
To estimate the rejection power for gamma-ray background as a function of 
energy from our data we proceed in three steps: First we split our data sample 
into energy bands according to the total p.e. count observed with both, PMT1 and 
PMT2. We then fit Gaussian distributions to both of the $R_{PSD}$ 
distributions, the one for nuclear recoils and the one for electron events, 
separately in each energy band. 
In the third step we estimate the contamination in the nuclear recoil sample 
from the tail of the gamma-ray Gaussian that extends beyond the mean of the 
nuclear recoil Gaussian. 
Using the fitted Gaussians was compared to counting events directly and 
was found to yield the same result. 
By integrating the tail of the Gaussian fitted to the electron event 
distribution that extends beyond the mean of the nuclear recoil $R_{PSD}$ 
distribution we define the fractional electron leakage $r_{EL}$ into our 
nuclear recoil sample at an efficiency of 50\% for nuclear recoils:
\begin{eqnarray}
r_{EL} = \int_{\mu_n}^{+\inf} G_e\,dx
\end{eqnarray}
Here $G_e$ denotes the normalized Gaussian as fitted to the $R_{PSD}$ 
distribution for electron events, which has a mean $\mu_e$ and a variance 
$\sigma_e$. 
$G_n$, $\mu_n$, and $\sigma_n$ stand for the corresponding entities derived 
from the fit to the $R_{PSD}$ distribution of the nuclear recoils. 

\begin{figure}[h]
\begin{center}
\includegraphics[scale = 0.5]{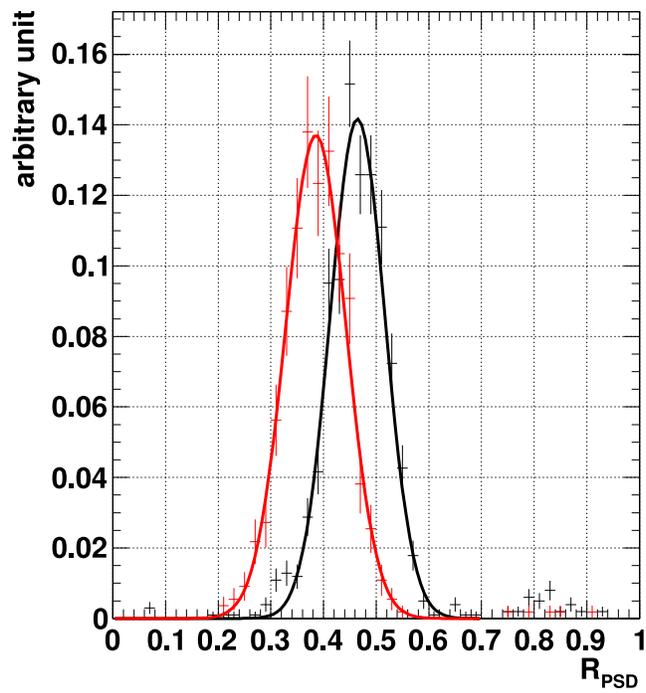}
\caption{The $R_{PSD}$ distribution in the energy range between 4.8-7.2\,keV$_{ee}$. The red and black lines show the $^{137}$Cs data and the $^{252}$Cf data, respectively.}
\label{fig:res1}
\end{center}
\end{figure}

 Electron leakage was evaluated for the p.e. ranges of 100$-$150, 150$-$200, 
200$-$250, 250$-$300, and 300$-$400\,p.e.  
 Using a conversion factor of 20.9\,p.e./keV$_{ee}$, these p.e. ranges correspond 
to energy ranges of 4.8$-$7.2, 7.2$-$9.6, 9.6$-$12, 12$-$14.4, and 14.4$-$19.1\,keV$_{ee}$ respectively. 
Fig.\ref{fig:res1} shows the $R_{PSD}$ distributions for both the nuclear and 
the electron event samples in the energy range from 4.8 to 7.2\,keV$_{ee}$. 
The corresponding Gaussian fits are also shown. 

\begin{figure}[h]
\begin{center}
\includegraphics[scale = 0.5]{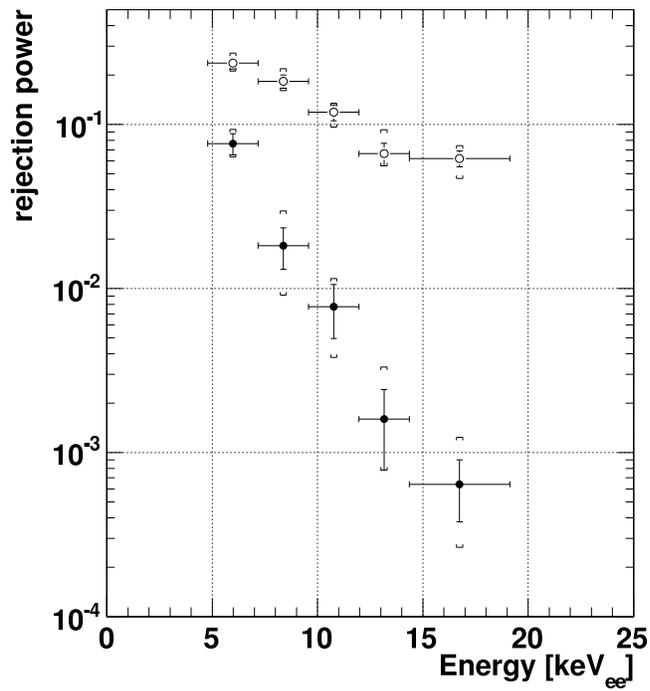}
\caption{The rejection power distribution. The open and filled circle points show the masked(4.6\,p.e./keV) and non-masked (20.9\,p.e./keV) data. Error bars shows the statistical error. Braces show the error including systematic uncertainty added in quadrature. 
}
\label{fig:re_sys2}
\end{center}
\end{figure}

Fig.\ref{fig:re_sys2} summarizes the fractional electron leakages we measured 
in the energy ranges listed above. 
Horizontal bars reflect the range of electron equivalent recoil energy 
that was used. 
Vertical error bars show the statistical uncertainties; the range between 
the braces has the systematic uncertainties added quadratically. 
The largest contribution to the systematic uncertainty came from 1\,p.e. determination. 
Open circles 
show the results for the study with the artificially reduced effective light 
yield of 4.6\,p.e./keV$_{ee}$, while solid circles show rejection power as a 
function of energy for the full measured scintillation light yield of 
20.9\,p.e./keV$_{ee}$. 

\begin{figure}[h]
\begin{center}
\includegraphics[scale = 0.5]{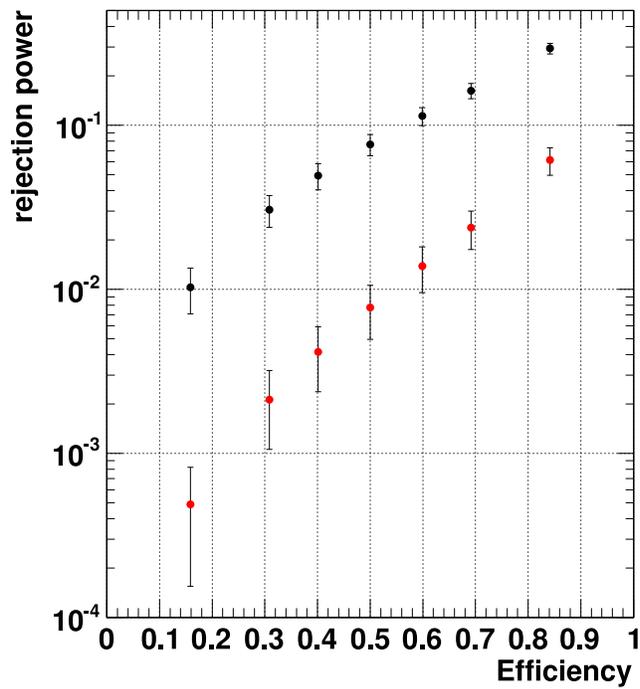}
\caption{The efficiency dependence of nuclear recoils retention for rejection power. The black and red points show the rejection power in the energy range 4.8-7.2 keV$_{ee}$ and 9.6-12 keV$_{ee}$ for non-masked data, respectively.}
\label{fig:eff}
\end{center}
\end{figure}

 The efficiency dependence of the nuclear recoils retention was also studied as shown in Fig.\ref{fig:eff}. The black and red points show the rejection power in the energy range 4.8-7.2 keV$_{ee}$ and 9.6-12 keV$_{ee}$ for non-masked data, respectively. The trade off between rejection of electron events and efficiency for nuclear recoils retention can be evaluated from this figure. 

\section{Discussion}

Our study clearly shows that PSD can be used in purely scintillation based 
LXe dark matter detectors. As expected its power depends on the effective 
light yield. Our $r_{EL}$ measure as defined above ultimately depends on the 
relative size of two quantities: the distance of the two means $\mu_n-\mu_e$ 
as compared to the width of the electron distribution $\sigma_e$. 

\begin{figure}[h]
\begin{center}
\includegraphics[scale = 0.5]{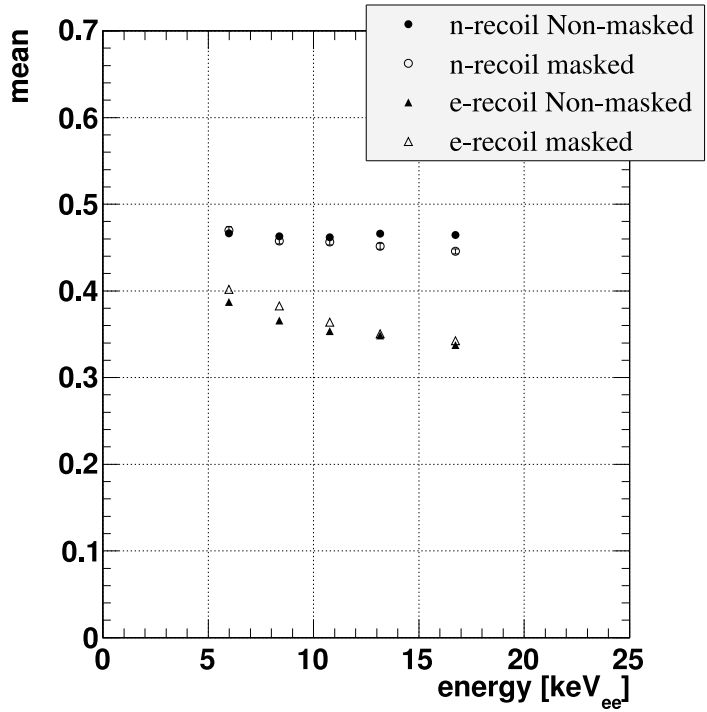}
\caption{Mean value of the $R_{PSD}$ as a function of observed energy in unit of keV$_{ee}$ for $^{137}$Cs runs and  $^{252}$Cf runs. The filled and open triangle points show the values of $^{137}$Cs runs for the non-masked PMT1+PMT2 (20.9\,p.e./keV) and for the masked PMT2 (4.6\,p.e./keV), respectively.
 The filled and open circle points show the values of $^{252}$Cf data for the non-masked PMT1+PMT2 (20.9\,p.e./keV) and for the masked PMT2 (4.6\,p.e./keV), respectively.}
\label{fig:v10_cs_m}
\end{center}
\end{figure}

In Fig.\ref{fig:v10_cs_m} we compare the mean values of the $R_{PSD}$ 
distributions for the electron and nuclear recoil runs as a function 
of our energy ranges. Both the high yield data and the data taken 
with an artificially lower effective yield are shown, and it can be seen that 
the means change by less than 5\% as we change the light yield. 
Yet the effect of this systematic change in $\mu_n-\mu_e$ contributes only 
4\% to the overall efficiency loss; the larger contribution comes from the 
widening on the distribution.

\begin{figure}[h]
\begin{center}
\includegraphics[scale = 0.5]{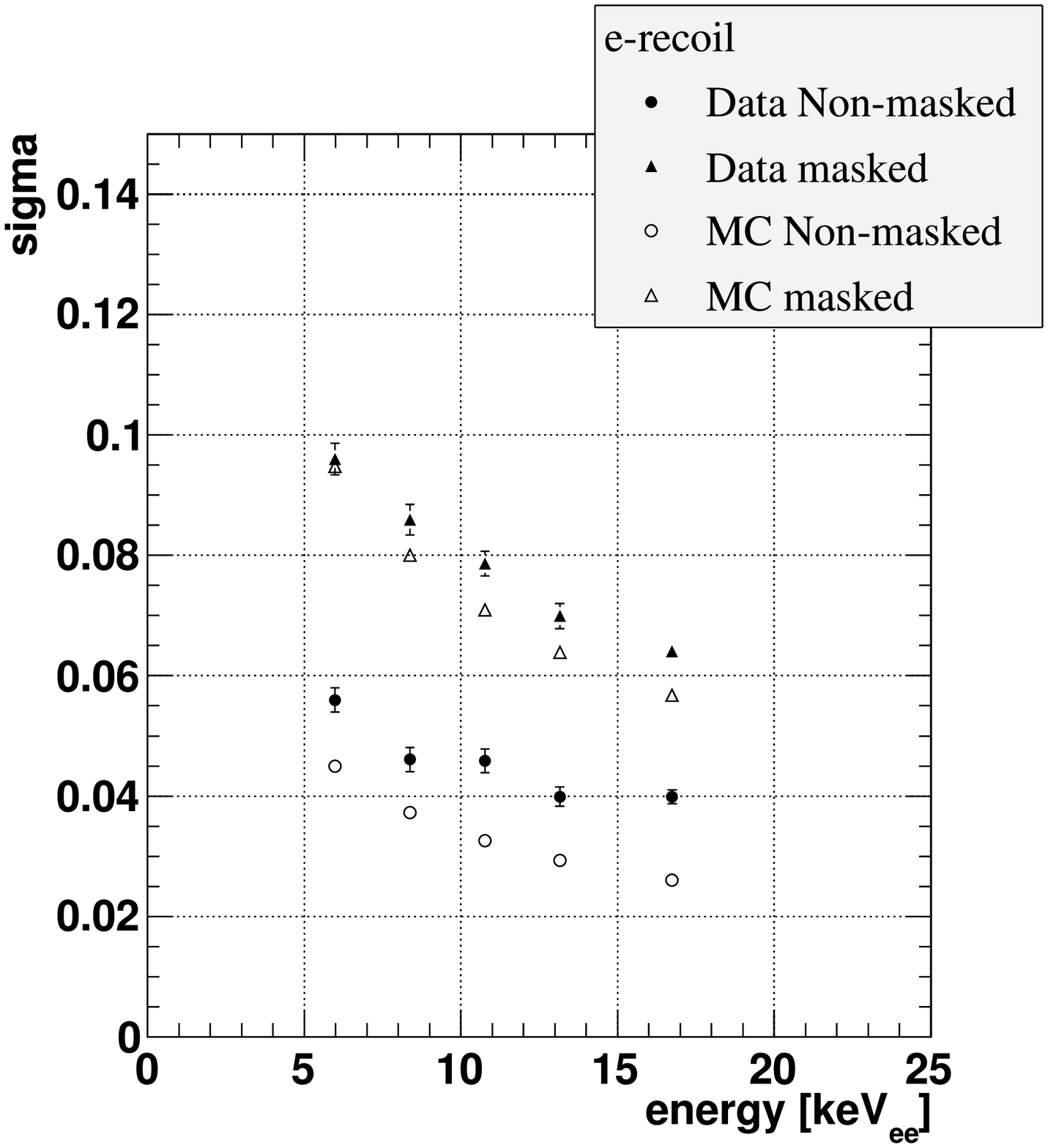}
\caption{Sigma of the $R_{PSD}$ distribution as a function of observed energy in unit of keV$_{ee}$ for $^{137}$Cs data. The filled circle and filled triangle points show the values for the non-masked PMT1+PMT2 data (20.9\,p.e./keV) and for the masked PMT2 data (4.6\,p.e./keV), respectively. The open circle and open triangle points show the MC simulation of the non-masked PMT1+PMT2 (20.9\,p.e./keV) and the masked PMT2 (light yield was reduced to 4.6\,p.e./keV), respectively.}
\label{fig:v10_cs_s}
\end{center}
\end{figure}

\begin{figure}[h]
\begin{center}
\includegraphics[scale = 0.5]{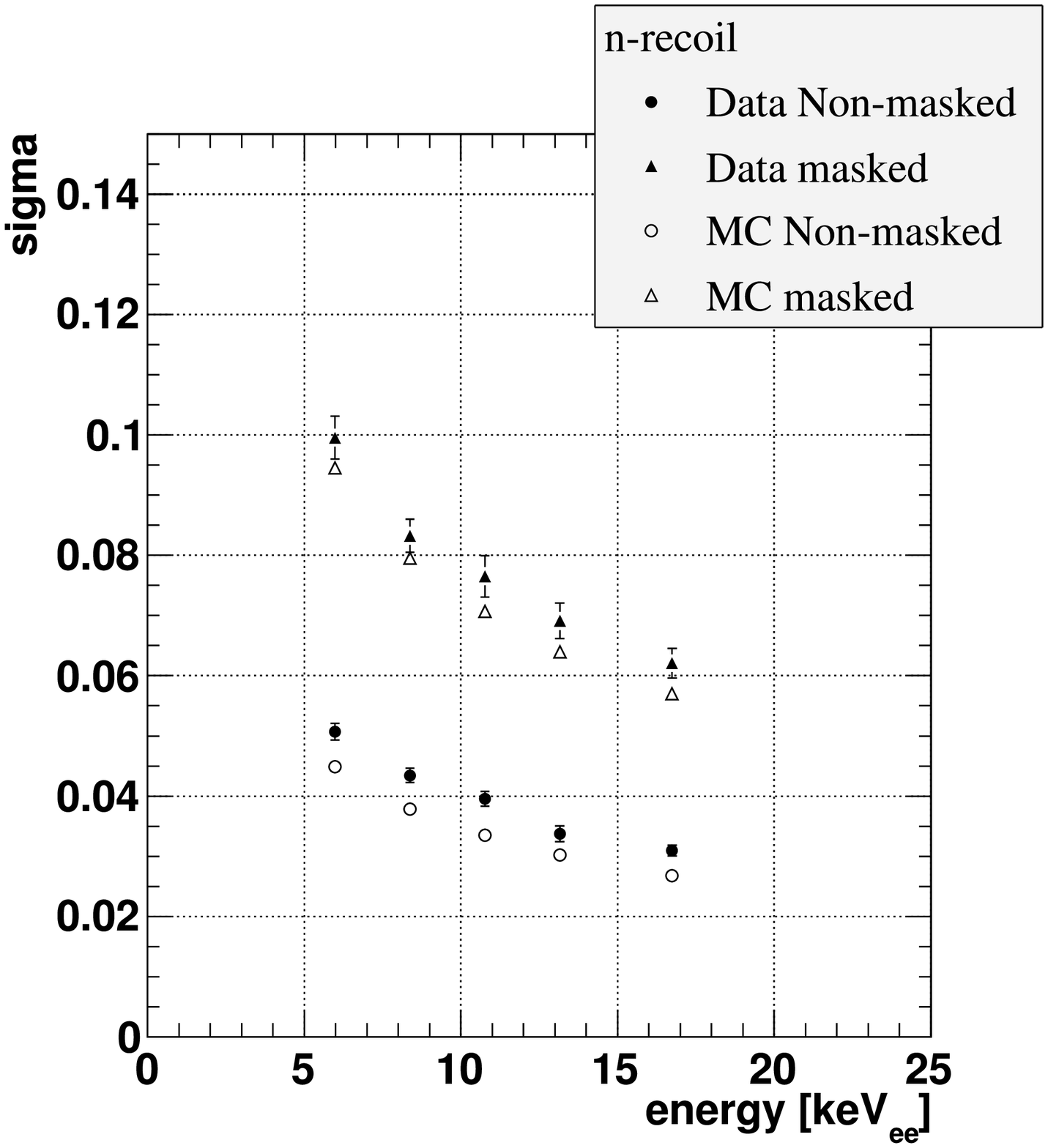}
\caption{Sigma of the $R_{PSD}$ distribution as a function of observed energy in unit of keV$_{ee}$ for $^{252}$Cf data. The filled circle and filled triangle points show the values for the non-masked PMT1+PMT2 (20.9\,p.e./keV) and for the masked PMT2 (4.6\,p.e./keV), respectively. The open circle and open triangle points shows the MC simulation of the non-masked PMT1+PMT2 (20.9\,p.e./keV) and the masked PMT2 (light yield was 4.6\,p.e./keV).}
\label{fig:v10_cf_s}
\end{center}
\end{figure}

Looking at the electron distribution width $\sigma_e$, we tried to separate 
the statistical 
component reflecting the p.e. statistics from an intrinsic component reflecting 
the physics of the various processes involved in generating, detecting, 
conditioning, and recording the signal. 
Fig.\ref{fig:v10_cs_s} and Fig.\ref{fig:v10_cf_s} respectively show $\sigma_e$ 
and $\sigma_n$, again as a function of our energy ranges.

\begin{figure}[h]
\begin{center}
\includegraphics[scale = 0.5]{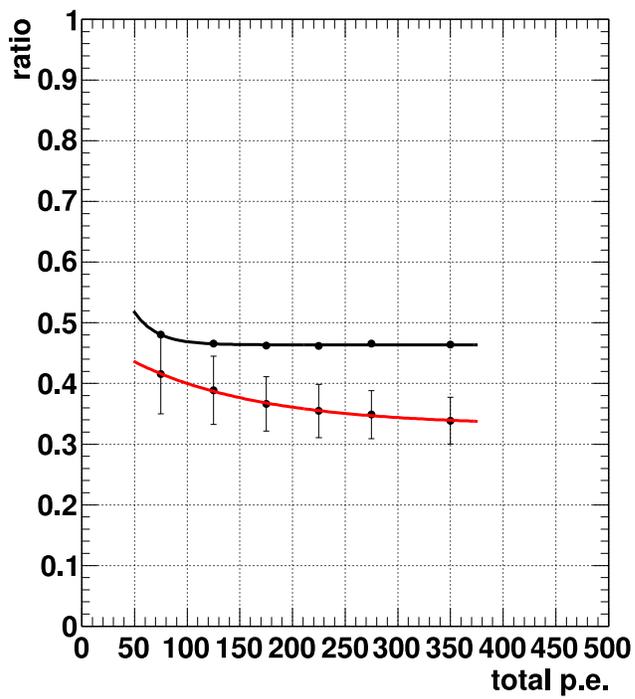}
\caption{The relation between $R_{PSD}$ and total p.e. The black and red lines show the fit function of $^{252}$Cf and $^{137}$Cs. The error bar of $^{137}$Cs represent the 1 sigma from the mean value.}
\label{fig:fit}
\end{center}
\end{figure}

To estimate the statistical contribution to the width of the $R_{PSD}$ Gaussians 
we used a Monte Carlo (MC) simulation. 
In this process we first randomly chose a value for $p.e._{tot}$ in the relevant 
range from 50 to 400\,p.e..  
Using the fits to our $R_{PSD}$ means as shown in Fig.\ref{fig:fit} we 
determine the proper $R_{PSD}$ mean to use for this $p.e._{tot}$ value. 
Using $p.e._{tot}$ and its proper $R_{PSD}$, the MC obtains a set of $p.e._{prmt}$ 
by sampling from the binomial distribution:
\begin{eqnarray}
 P(p.e._{prmt}) = _{p.e._{tot}}C_{p.e._{prmt}}  R_{PSD}^{p.e._{prmt}} (1-R_{PSD})^{p.e._{tot}-p.e._{prmt}}   
\end{eqnarray}  
From this set of $p.e._{prmt,MC}$ we calculate a set of 
$R_{PSD,MC}=p.e._{prmt,MC}/p.e._{tot}$.  $C$ means binominal coefficient. 

$R_{PSD,MC}$ distributions are built up for both the electron and nuclear recoil 
means and fitted to Gaussians to extract $\mu_{e,MC}$, $\sigma_{e,MC}$, 
$\mu_{n,MC}$ and $\sigma_{n,MC}$. 
The resulting $\mu_{e,MC}$ and $\mu_{n,MC}$ are in good agreement with their 
measured counterparts $\mu_e$ and $\mu_n$. 

Fig.\ref{fig:v10_cs_s} and Fig.\ref{fig:v10_cf_s} show the variances of $\sigma$ for MC and 
data compared for electron and nuclear recoil respectively. Systematically the 
MC data have less spread than the real data, but the effect is quite striking 
for the high light yield electron event data. 

\begin{figure}[h]
\begin{center}
\includegraphics[scale = 0.5]{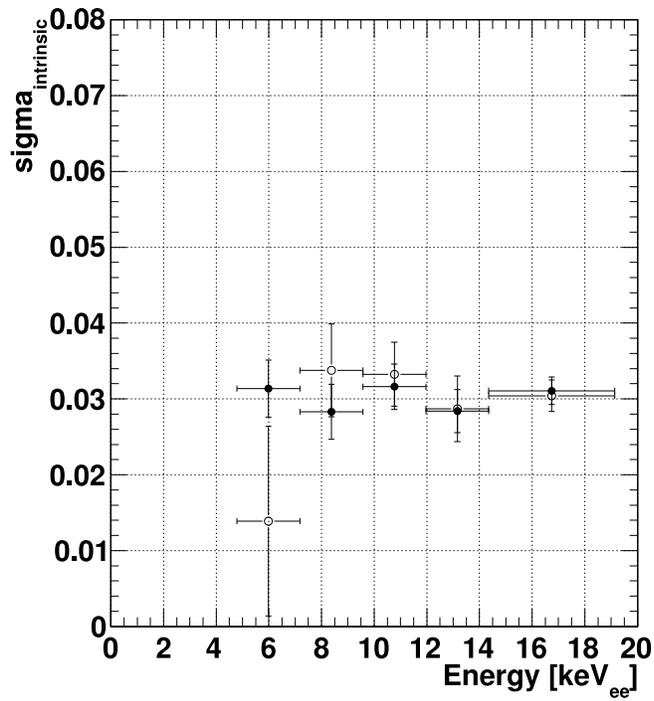}
\caption{The $\sigma_{intrinsic}$ distribution of electron events. The open and filled circle points show the values of the masked and the non-masked data, respectively.}
\label{fig:unknown}
\end{center}
\end{figure}

 In the case of electron event the waveform is dominated by recombination, which 
is slow. In case of nuclear recoil, the waveform reflects the relatively short 
lifetimes of the singlet and triplet states. 
Fluctuations in the energy deposit along the electron track will translate into 
fluctuations in the
recombination time scale and might be one mechanism by which additional 
fluctuations can propagate into our $R_{PSD}$ parameter. 
In an effort to quantify the extra contribution $\sigma_{intrinsic}$ to the width 
of our distribution that is evident in our data we calculated:
\begin{eqnarray}
\sigma_{intrinsic} = \sqrt{\sigma^2_e - \sigma^2_{e,MC}}
\end{eqnarray}
The results are summarized in Fig. \ref{fig:unknown}. The observed consistency 
in all over energy range gives support to the 
hypothesis that we are seeing the effects of processes at the light emission 
stage, as that part was not changed; our light yield was only effectively 
reduced by a mask. Intrinsically the light yield in the low and high yield data runs was still the same. 

\section{Conclusion}

 PSD in liquid xenon was studied at energies relevant to dark matter searches.
A significant difference in pulse shape between nuclear recoil and electron events was exploited in a high light yield setup in which 5\,keV$_{ee}$ 
energy deposit produce 100\,p.e. in PMT signal output. 

 At high light yield (20.9\,p.e./keV) our $R_{PSD}$ parameter allows a rejection 
of electron events to a level of 
7.7$\pm$1.1(stat)$\pm ^{1.2}_{0.6}$(sys)$\times$10$^{-2}$ in the energy range 
between 4.8-7.2\,keV$_{ee}$ with a 50\% efficiency to retain nuclear recoil 
events.
 In the energy range 14.4-19.1\,keV$_{ee}$ electron events were reduced by more than 3 orders of magnitude with the same efficiency for nuclear recoils. 

 The dependence of this rejection power on photon statistics was also studied.
At low effective light yield (4.6\,p.e./keV), a rejection of electron events to 
2.4$\pm$0.2(stat)$\pm ^{0.3}_{0.2}$(sys)$\times$10$^{-1}$ was demonstrated for the 
energy range 4.8-7.2\,keV$_{ee}$ with a 50\% efficiency for nuclear recoil events.

 In our MC replication of the experimental results the width of the $R_{PSD}$ distribution for nuclear recoils is almost exhausted by the expected purely statistical contribution to that width, while that is clearly not the case for electron events. This excess width was tentatively interpreted as stemming from fluctuations 
inherent to the energy deposit of electron events.  

\section{Acknowledgements}
 We gratefully acknowledge the cooperation of Kamioka Mining and Smelting Company.
 This work was supported by the Japanese Ministry of Education, Culture, Sports, Science and Technology, and Grant-in-Aid for Scientific Research.
 We are supported by Japan Society for the Promotion of Science.

% Figure

% Figure

% Figure

% Figure

% Figure

% Figure

% Figure

% Figure

% Figure

% Figure

% Figure

% Figure

% Figure

\end{document}